\documentclass[aps,prb,twocolumn,showpacs]{revtex4}
\usepackage{mathptmx}
\usepackage{graphicx}
\usepackage{bm}
\def \virg{,}
\def \point{.}
\def \bk{\mathbf{k}}
\def \bq{\mathbf{q}}
\def \e-h{$e$-$h$}
\renewcommand{\Im}{\mathop{\rm Im}\nolimits}
\begin{document}
\title{Excitonic BCS-BEC crossover at finite temperature: 
Effects of repulsion and electron-hole mass difference}

\author{Yuh Tomio}%
\author{Kotaro Honda}%
\author{Tetsuo Ogawa}%
\affiliation{CREST, JST, and 
Department of Physics, Osaka University, 
Toyonaka, Osaka 560-0043, Japan}%

\date{\today}

\begin{abstract}
The BCS to Bose-Einstein condensation (BEC) crossover 
  of electron-hole (\e-h) pairs
  in optically excited semiconductors is studied using
  the two-band Hubbard model with both repulsive 
  and attractive interactions.
Applying the self-consistent $t$-matrix approximation combined with 
  a local approximation, we examine the properties of 
  a normal phase and an excitonic instability.
The transition temperature from the normal phase
  to an \e-h pair condensed one is studied to clarify  
  the crossover from an \e-h BCS-like state to an excitonic 
  Bose-Einstein condensation, which takes place on increasing 
  the \e-h attraction strength. 
To investigate effects of the repulsive interaction 
  and the \e-h mass difference,
  we calculate the transition temperature for various 
  parameters of the interaction strengths, the \e-h particle density, 
  and the mass difference.
While the transition temperature in the \e-h BCS regime is 
  sufficiently suppressed by the repulsive interaction, 
  that of the excitonic BEC is largely insensitive to it.
We also show quantitatively that in the whole regime 
  the mass difference leads to 
  large suppression of the transition temperature.
\end{abstract}

\maketitle

\section{Introduction}

The Bose-Einstein condensation (BEC) of excitons in solids 
  is one of the most interesting subjects 
  in condensed matter physics.%
~\cite{Moskalenko00}
In this phenomenon the attractive Coulomb interaction between 
  electrons and holes plays an essential role 
  in forming bosonic electron-hole (\e-h) bound pairs. 
This interaction and the bosonic nature strongly affect 
  the statistical and thermodynamic properties of \e-h systems.

In \e-h systems realized in photoexcited semiconductors,    
  therefore, various remarkable properties are expected 
  to depend on \e-h density, temperature, etc., 
  and they have been investigated extensively both 
  experimentally and theoretically.%
~\cite{Haug84,Zimmermann}
In particular, metal-insulator transitions 
  have attracted interest for many years: 
  One is the exciton Mott transition from 
  an exciton or biexciton insulating gas phase to 
  an \e-h plasma (normal) metallic phase with increasing \e-h density. 
Another, which is our main target in this work, is  
  the transition from the normal phase to an \e-h pair condensed one. 
At low \e-h density, strongly bound \e-h pairs are 
  expected to undergo BEC as an exciton gas 
  at cryogenic temperatures.%
~\cite{Moskalenko62,Blatt62}
On the other hand, at high \e-h density 
  where the mean interparticle distance
  is shorter than the exciton Bohr radius, 
  weakly bound \e-h pairs should behave like the Cooper pairs 
  in conventional superconductors 
  at sufficiently low temperatures, 
  that is, the Bardeen-Cooper-Schrieffer (BCS) state of \e-h pairs.%
~\cite{Keldysh65,Jerome67}  
There have been many experimental attempts to observe 
  such a coherent condensation of \e-h bound pairs in solids
  such as Cu$_2$O, CuCl and GaAs.%
~\cite{Snoke03,Snoke90,Lin93,Gonokami02,Vasilev04} 
At present, however, definitive evidence is still lacking in experiments.

Since the \e-h recombination 
  is usually much slower than the intraband relaxation,   
  we focus on a quasi-thermal-equilibrium state of the \e-h system. 
In this situation, the crossover problem%
~\cite{Comte82,InagakiSSC,Inagaki02,Littlewood04} 
  between the BEC of excitons and the \e-h BCS-like state is fascinating,   
  especially from the viewpoint of the difference from the BCS-BEC 
  crossover in superconductors or trapped atomic Fermi gases.%
~\cite{Nozieres85,Micnas90,Randeria,Ohashi02} 
We believe that revealing such differences will 
  deepen our understanding of \e-h systems,  
  leading toward the experimental observation. 
Compared with other systems undergoing condensation of bound pairs,  
  electrons and holes in a quasi-thermal-equilibrium system 
  in semiconductors have the following two notable characteristics: 
(i)~They involve not only the \e-h attractive Coulomb interaction 
  but also the {\it repulsive} one between like particles 
  (besides the Pauli exclusion principle), and  
(ii)~they generally have different masses and mass anisotropies.
To understand the coherent condensation of \e-h pairs, therefore, 
  we highlight these important characteristics in this work. 
Early and quite recent works%
~\cite{Kopaev66,Zittartz67,Mizoo05}  
  based only on the BCS-like mean-field theory have shown that 
  the mass difference and the mass anisotropy suppress 
  the \e-h BCS order. 
In the presence of such asymmetry, however, 
  it is still unclear how the crossover from the \e-h BCS state to 
  the excitonic BEC evolves. 
In addition, little attention has been paid to 
  the roles of electron-electron ($e$-$e$) and hole-hole ($h$-$h$) 
  repulsive interactions in this problem.

In this paper, 
  by calculating the transition temperature $T_c$ that 
  directly reflects the excitonic BCS-BEC crossover, 
  we clarify the effects of the repulsive interaction and 
  the mass difference on the \e-h pair condensation 
  from the BCS to the BEC regime. 
A simple two-band Hubbard model with attractive and repulsive 
  on-site interactions is adopted to describe the \e-h systems. 
Here we suppose that conduction electrons and valence holes, 
  whose bands are isotropic, have infinite lifetime, and 
  the number of electrons is equal to that of holes ($N_e=N_h$).
In our model the interaction strengths and the \e-h density are 
  treated as independent parameters.
Several physical quantities, such as the density of states, 
  the density of occupied sites, and the quasiparticle weights,  
  are calculated to discuss the properties of the normal phase.

We employ the self-consistent $t$-matrix approximation (SCTMA) 
  in our analysis. The SCTMA is an effective method for 
  the BCS-BEC crossover problem at finite temperatures,%
~\cite{Fresard92,Haussmann,Keller99}  
  and is a conserving approximation based on the Baym-Kadanoff theory.
This approximation deals correctly with two-particle correlations.  
Thus it becomes asymptotically exact in the low-density limit.
In addition to the SCTMA, we also use a local approximation (LA),   
  which is justified in high spatial dimensions. 
The procedure is to neglect the momentum dependence of 
  the self-energy and the vertex function.%
~\cite{Keller99}   
It is in the same spirit as the $\bk$-averaged approximation%
~\cite{Letz98} 
  or the dynamical mean-field theory (DMFT).%
~\cite{Georges96}  
As known in the DMFT literature, 
  the LA itself (without other approximations)
  becomes exact in the limit of infinite spatial dimensions 
  and a good approximation for three-dimensional systems. 
Although the SCTMA combined with the LA is not exact 
  even in the infinite-coordination limit since 
  only ladder diagrams are summed up, 
  it has been successful for the superconductivity of 
  the {\it single-band} attractive Hubbard model in high dimensions.%
~\cite{Keller99}
In particular,
  the successive interpolation between the BCS limit  
  with $T_c \propto \exp(t/U')$ and the BEC limit with 
  $T_c \propto t^2/U'$ can be described well,%
~\cite{Nozieres85,Micnas90}
  where $U'$ and $t$ denote the attractive interaction 
  and the transfer energy, respectively. 
Therefore, we extend the scheme to our two-band model, and  
  expect that our results are valid for three-dimensional bulk systems.

This paper is organized as follows. 
In Sec.~II, the SCTMA combined with the LA is 
  applied to the normal phase for the \e-h two-band Hubbard model. 
In Sec.~III, several physical quantities are calculated and 
  properties of the normal phase are discussed. 
In Sec.~IV, the transition temperature from the normal phase 
  to the \e-h pair condensed phase is presented 
  as a function of the \e-h attraction strength for various parameters. 
We examine in Sec.~IV~A the effect of the $e$-$e$ ($h$-$h$)
  repulsion on the transition temperature.  
In Sec.~IV~B, the effect of the mass difference on 
  the transition temperature is analyzed and 
  features of the BCS-BEC crossover in the \e-h system are discussed. 
Concluding remarks are given in Sec.~V.

\section{Formulation}

We consider an \e-h system described by 
  the two-band Hubbard model. 
The Hamiltonian is given by 
\begin{eqnarray}
 \label{H}
 H &=& -\sum_{\langle ij \rangle,\sigma}\sum_{\alpha=e,h}
     t_\alpha c_{i\sigma}^{\alpha\dagger} c_{j\sigma}^\alpha
    -  \sum_{j\sigma,\alpha} \mu_\alpha
n_{j\sigma}^\alpha
\nonumber \\
&& {}
 + U \sum_{j,\alpha} 
n_{j\uparrow}^\alpha n_{j\downarrow}^\alpha 
- U' \sum_{j\sigma\sigma'} 
n_{j\sigma}^e n_{j\sigma'}^h
\virg 
\end{eqnarray}
where $c^{e\dagger}_{j\sigma}$ ($c^{h\dagger}_{j\sigma}$) 
  denotes a creation operator of an electron (a hole) 
  with spin $\sigma=\{\uparrow,\downarrow\}$ at the $j$th site
  and $n_{j\sigma}^\alpha=c_{j\sigma}^{\alpha\dagger} c_{j\sigma}^\alpha$ 
  with $\alpha=\{e,h\}$. 
The quantities $t_e$ ($t_h$) and $\mu_e$ ($\mu_h$) are  
  the transfer integral of the electrons (holes) 
  between the nearest-neighbor sites and 
  the chemical potential measured from the center of 
  the bare electron (hole) band, respectively. 
The on-site Coulomb interaction 
  of the $e$-$e$ ($h$-$h$) repulsion and 
  that of the \e-h attraction 
  are expressed by $U$ and $-U'$, respectively.

\begin{figure}[tb]
 \includegraphics[width=8cm,clip,keepaspectratio]{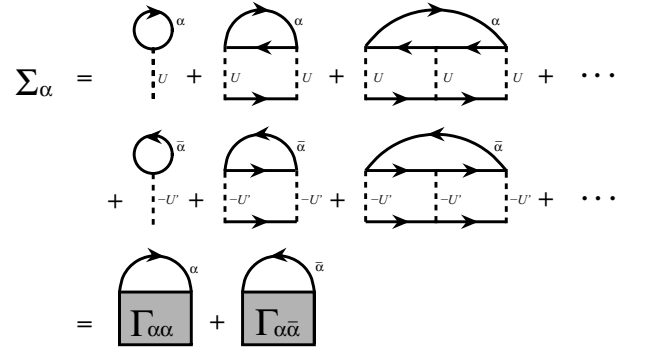}
\caption{%
Feynman diagrams for the self-energy in the normal phase.
The solid line denotes the electron or hole Green's function 
  $G_\alpha$ where $\bar{\alpha}=h$ ($e$) for $\alpha=e$ ($h$). 
The spin weight is 2 for the diagram of the vertex function 
  $\Gamma_{\alpha\bar{\alpha}}$.
}%
\label{Fig1}
\end{figure}

We apply the SCTMA to the model~(\ref{H}). 
Feynman diagrams contributing to the self-energy 
  of electrons and holes in the normal phase within the SCTMA 
  are shown  in Fig.~1, where all particle-hole (particle-particle) 
  ladder diagrams are taken into account with respect to 
  the interaction $U$ ($-U'$). 
The explicit expression of the self-energy is given by 
\begin{eqnarray}
\label{Sigma}
 \Sigma_\alpha (\bk,\omega_n) &=& 
\frac{T}{N} \sum_{\bq,\nu_m} e^{i(\nu_m+\omega_n)0^+}
\Gamma_{\alpha\alpha}(\bq,\nu_m) 
\nonumber \\
&& {} 
\times G_\alpha(\bq+\bk,\nu_m+\omega_n)
\nonumber \\
&& {}
+\frac{2T}{N} \sum_{\bq,\nu_m} e^{i(\nu_m-\omega_n)0^+}
\Gamma_{\alpha\bar{\alpha}}(\bq,\nu_m) 
\nonumber \\
&& {} 
\times G_{\bar{\alpha}}(\bq-\bk,\nu_m-\omega_n)
\virg 
\end{eqnarray}
where $N$ is the total number of lattice sites, 
  $T$ denotes the temperature, 
  $\omega_n=(2n+1)\pi T$, and $\nu_m=2\pi m T$ 
  with integer $n$ and $m$.
The symbol $0^+$ in the convergence factor denotes 
  a time infinitesimally later than $\tau=0$. 
  Here the spin index $\sigma$ is omitted because we now consider 
  the spin-symmetric case, but the spin weight (a factor 2) is counted. 
The single-particle Matsubara Green's function 
  $G_\alpha(\bk,\omega_n)$ defined by the Fourier transform of 
  $-\langle T_\tau c_{j\sigma}^\alpha(\tau) 
  c_{j'\sigma}^{\alpha\dagger} \rangle$
  is expressed in terms of the self-energy~(\ref{Sigma}) as 
\begin{equation}
 \label{G}
G_\alpha(\bk,\omega_n)= \frac{1}
{i\omega_n+\mu_\alpha-\epsilon^\alpha_\bk-\Sigma_\alpha(\bk,\omega_n)}
\virg
\end{equation}
where $\epsilon^\alpha_\bk$ is the band dispersion of 
  the noninteracting electrons/holes.
The two-particle vertex functions $\Gamma_{\alpha\alpha'}(\bq,\nu_m)$
  are obtained as the infinite geometric series, 
\begin{eqnarray}
 \label{Gamma_aa}
\Gamma_{\alpha\alpha}(\bq,\nu_m) &=& 
U + U K_{\alpha\alpha}(\bq,\nu_m) U + \cdots
\nonumber \\
&=& \frac{U}{1-U K_{\alpha\alpha}(\bq,\nu_m)}
\virg \\
 \label{Gamma_eh}
\Gamma_{\alpha\bar{\alpha}}(\bq,\nu_m) &=& 
-U' + (-U') K_{\alpha\bar{\alpha}}(\bq,\nu_m) (-U') + \cdots
\nonumber \\
&=& \frac{-U'}{1+U' K_{\alpha\bar{\alpha}}(\bq,\nu_m)}
\virg 
\end{eqnarray}
where the pair propagators are 
\begin{eqnarray}
 \label{K_aa}
K_{\alpha\alpha}(\bq,\nu_m) &=& 
-\frac{T}{N}\sum_{\bk,\omega_n} 
G_\alpha(\bk,\omega_n) G_\alpha(\bq+\bk,\nu_m+\omega_n)
\virg~~~  
\\
 \label{K_eh}
K_{\alpha\bar{\alpha}}(\bq,\nu_m) &=& 
-\frac{T}{N}\sum_{\bk,\omega_n} 
G_\alpha(\bk,\omega_n) G_{\bar{\alpha}}(\bq-\bk,\nu_m-\omega_n)
\point~~~  
\end{eqnarray}

By using the self-consistent set of Eqs.~(\ref{Sigma})-(\ref{K_eh}), 
  the model~(\ref{H}) can be solved within the SCTMA. 
However, a straightforward calculation is very difficult 
  especially for three-dimensional systems with large system size $N$. 
In this work, then, we combine a LA 
  with the SCTMA to simplify the above self-consistent calculation.%
~\cite{Letz98,Keller99}   
The procedure is performed by neglecting 
  momentum dependence of the self-energy and the vertex function  
  [i.e., $\Sigma_\alpha(\bk,\omega_n) \rightarrow
  \Sigma_\alpha(\omega_n)$ 
  and  
  $\Gamma_{\alpha\alpha'}(\bq,\nu_m) \rightarrow 
  \Gamma_{\alpha\alpha'}(\nu_m)$].
This analysis is equivalent to solving an effective single-impurity
  problem of the DMFT within the SCTMA. 
By making the LA, 
  the self-consistent Eqs.~(\ref{Sigma})-(\ref{K_eh}) can be written  
  in terms of only local functions: 
\begin{eqnarray}
\label{Sigma_local}
 \Sigma_\alpha (\omega_n) &=& 
T \sum_{\nu_m} e^{i(\nu_m+\omega_n)0^+}
\Gamma_{\alpha\alpha}(\nu_m) G_\alpha(\nu_m+\omega_n)
\nonumber \\
&& {}
+ 2T \sum_{\nu_m} e^{i(\nu_m-\omega_n)0^+}
\Gamma_{\alpha\bar{\alpha}}(\nu_m) G_{\bar{\alpha}}(\nu_m-\omega_n)
\virg~~~ 
\end{eqnarray}
where the local Green's function is 
\begin{eqnarray}
 \label{G_local}
G_\alpha(\omega_n) &=& 
\frac{1}{N} \sum_{\bk} G_\alpha(\bk,\omega_n)
\nonumber \\
&=& 
\int d\epsilon 
\frac{\rho^0_\alpha(\epsilon)}
{i\omega_n+\mu_\alpha-\epsilon-\Sigma_\alpha(\omega_n)}
\virg
\end{eqnarray}
with the noninteracting density of states $\rho_\alpha^0(\epsilon)$, 
  and the local vertex functions are 
\begin{eqnarray}
 \label{Gamma_aa_local}
\Gamma_{\alpha\alpha}(\nu_m) &=& 
\frac{U}{1-U K_{\alpha\alpha}(\nu_m)}
\virg \\
 \label{Gamma_eh_local}
\Gamma_{\alpha\bar{\alpha}}(\nu_m) &=& 
\frac{-U'}{1+U' K_{\alpha\bar{\alpha}}(\nu_m)}
\virg 
\end{eqnarray}
with the local pair propagators 
\begin{eqnarray}
 \label{K_aa_local}
K_{\alpha\alpha}(\nu_m) &=& 
-T \sum_{\omega_n} G_\alpha(\omega_n) G_\alpha(\nu_m+\omega_n)
\virg  \\
 \label{K_eh_local}
K_{\alpha\bar{\alpha}}(\nu_m) &=& 
-T \sum_{\omega_n} G_\alpha(\omega_n) G_{\bar{\alpha}}(\nu_m-\omega_n)
\point
\end{eqnarray}

To determine the transition temperature of the condensation of \e-h pairs, 
  we examine the \e-h singlet pair susceptibility%
~\cite{Note0}
  $\chi(\bq,\nu_m)$, defined by the Fourier transform of 
  $-\langle T_\tau c_{j,-\sigma}^h(\tau) c_{j\sigma}^e(\tau)
  c_{j'\sigma}^{e\dagger}(\tau') c_{j',-\sigma}^{h\dagger}(\tau') \rangle$, 
  which is given by 
  the ladder terms corresponding to $\Gamma_{eh}(\bq,\nu_m)$, i.e., 
\begin{equation}
\label{chi}
 \chi(\bq,\nu_m) = \frac{K_{eh}(\bq,\nu_m)}{1+U'K_{eh}(\bq,\nu_m)}
\point
\end{equation}
Here note that within the present SCTMA the repulsive interaction $U$ 
  does not appear explicitly in Eq.~(\ref{chi}), but it influences 
  the \e-h pair susceptibility through the single-particle 
  Green's functions in Eq.~(\ref{K_eh}).  
If the uniform static \e-h pair susceptibility $\chi(\mathbf{0},0)$ 
  diverges for $T \searrow T_c$, 
  it is a signal of the onset of the \e-h pair condensation 
  (the Thouless criterion). 
Thus the transition temperature $T_c$ can be determined 
as the temperature satisfying the condition   
\begin{equation}
\label{Tc}
1+U'K_{eh}(\mathbf{0},0) = 0
\point
\end{equation}
Since the self-energy is local (i.e., momentum independent), 
  the pair propagator $K_{eh}(\mathbf{0},0)$ 
  in Eq.~(\ref{Tc}) can be expressed in terms of 
  only local functions as 
\begin{eqnarray}
\label{K_eh00}
K_{eh}(\mathbf{0},0) &=& 
-\frac{T}{N}\sum_{\bk,\omega_n} 
G_e(\bk,\omega_n) G_h(-\bk,-\omega_n)
\nonumber \\
&=& {} -T \sum_{\omega_n} 
\frac{G_e(\omega_n)-\gamma G_h(\omega_n)^*}
{\zeta_h(\omega_n)^*-\gamma \zeta_e(\omega_n)}
\virg 
\end{eqnarray}
where $\zeta_\alpha(\omega_n)=i\omega_n+\mu_\alpha-\Sigma_\alpha(\omega_n)$
  and $\gamma=t_h/t_e$ is defined as the \e-h band mass ratio $m_e/m_h$. 
Here we have assumed the relation
  $\epsilon^h_{\bk}=\gamma\epsilon^e_{\bk}$. 
In this paper we use 
  $\rho^0_\alpha(\varepsilon)=
  \sqrt{4t_\alpha^2-\varepsilon^2}/(2\pi t_\alpha^2)$ 
  as a typical density of states of three-dimensional systems.

\section{Properties of Normal Phase}

In this section, 
  properties of the normal phase above $T_c$ are discussed  
  by using the self-consistent solution of 
  Eqs.~(\ref{Sigma_local})-(\ref{K_eh_local}).  
We analyze the interacting density of states, 
  the density of sites occupied by electrons and holes, 
  and the quasiparticle weight, 
  for various values of $U$, $U'$, $\gamma=t_h/t_e$,  
  and the particle density $n$.  
In the following we consider the cases of $n=0.25$ and 0.5 
  (and $n=1$ only in the next section), where 
  the particle density is defined as 
\begin{equation}
\label{n}
n \equiv 
\sum_\sigma \langle  n_{j\sigma}^{\alpha} \rangle
= 2 T \sum_{\omega_n} e^{i\omega_n 0^+} G_\alpha(\omega_n) 
\virg
\end{equation}
under the condition of charge neutrality 
  $\left<n_{\sigma}^e\right>=\left<n_{\sigma}^h\right>$. 
Throughout the paper the quantity $t_e+t_h$ is taken as the energy unit.

\subsection{Density of states}

\begin{figure}[tb]
 \includegraphics[width=8cm,clip,keepaspectratio]{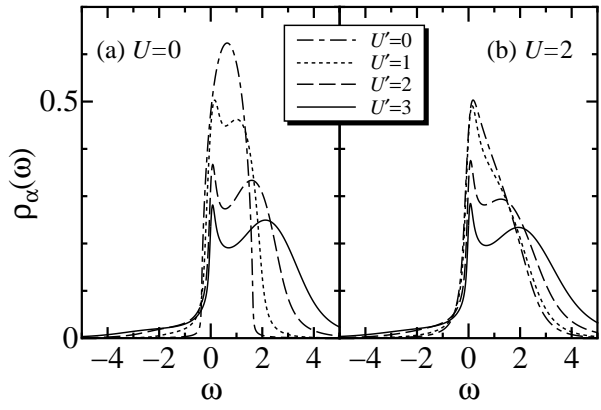}
\caption{%
The density of states $\rho_\alpha(\omega)$
  for $n=0.25$, $T=0.04$, and $\gamma=t_h/t_e=1$. 
All the energies are scaled by $t_e+t_h$.
}%
\label{Fig2}
\end{figure}

Figure~\ref{Fig2} shows the interacting density of states per spin, 
\begin{equation}
\label{rho}
\rho_\alpha(\omega)=-\frac{1}{\pi}\Im G_\alpha(\omega+i0^+)
\virg
\end{equation}
for various values of $U'$.
Parameters are chosen as $n=0.25$ (1/8 filling), 
  $T=0.04$ (that is, above $T_c$), and $\gamma=1$ 
  with the fixed  $U=$ (a) 0 and (b) 2.
Note that $\rho_e(\omega)=\rho_h(\omega)$ for $\gamma=1$.
  Here we have obtained the retarded Green's function 
  $G_\alpha(\omega+i0^+)$ 
  from the numerical analytic continuation 
  by using the Pad\'e approximation.

When $U=0$, $\rho_\alpha(\omega)$ exhibits a characteristic 
  two-peak structure as $U'$ is increased:  
  a sharp peak 
  (i.e., the quasiparticle coherent peak) appears 
  at $\omega \simeq 0$ and its weight decreases, 
  while a broad incoherent peak develops at $\omega \simeq U'$.
We should remark that in the low-density limit ($n \to 0$)  
  an \e-h bound state is formed for $U' \geq 1$ 
  and the binding energy is given by $E_b \simeq U'$. 
Hence, such a behavior of the density of states implies that 
  local \e-h pairs (excitons) tend to be formed,  
  which will become clear also from the analysis of the density of
  occupied sites (Fig.~\ref{Fig3}).

When $U=2$, $\rho_\alpha(\omega)$ is broadened by $U$,  
  so it is already strongly renormalized even for $U'=0$.
This indicates that $U$ gives the dominant contribution to 
  the band gap renormalization. 
For $U' < U$ ($U'=1$), in contrast to the case of $U=0$, 
  the broad incoherent peak is not observed. 
This means that the excitonic correlation is suppressed by $U$. 
For $U' \gtrsim U$ ($U'=2$ and 3),  
  $\rho_\alpha(\omega)$ shows the two-peak structure again. 
Compared with the case of $U=0$, however, the coherent (incoherent) 
  peak is enhanced (suppressed) especially for $U \simeq U'$. 
This is due to the competition between $U$ and $U'$. 
Similar behavior has been found more prominently 
  in the DMFT for $n=1$ (half filling).%
~\cite{Koga02,Tomio05}  
There, the metallic phase is stabilized for $U \simeq U' \leq 5$
  between the Mott-Hubbard and biexcitonlike insulating phases.   
Except in such a case,  
  the density of states in Fig.~\ref{Fig2} suggests fundamentally that 
  the metallic character tends to be lost because of both 
  the weight shifting due to $U'$ and the overall broadening due to $U$.

Here we note that 
  within the conserving approximation such as the present SCTMA, 
  the system never becomes an insulator
  [$\rho_\alpha(0)=0$] with no broken symmetry
  for any values of $U$ and $U'$.  
This is a well-known failure of this scheme.%
~\cite{Menge91,Keller99}

\subsection{Density of occupied sites}

Examining the density of occupied sites is an effective way to see 
  how the excitonic correlation increases in the normal phase. 
Now we focus on  
  the density of sites quadruply occupied by electrons and holes, 
  $D_{eh} \equiv \sum_{\sigma\sigma'}
  \langle n_{j\sigma}^{e} n_{j\sigma'}^{h} \rangle$. 
This is local susceptibility at equal times 
  and can be written in terms of 
  the local pair propagator (\ref{K_eh_local}): 
\begin{equation}
D_{eh} = 
\sum_{\sigma\sigma'} \langle n_{j\sigma}^{e} n_{j\sigma'}^{h} \rangle = 
-T \sum_{\nu_m} e^{i\nu_m 0^+}
\frac{4K_{eh}(\nu_m)}{1+U'K_{eh}(\nu_m)}
\point 
\end{equation}
We first note the limiting cases. 
For the noninteracting case $D_{eh}=n^2$. 
If all electrons and holes are perfectly bound 
  as local \e-h pairs by $U'$,  
  the density of occupied sites is expected to become 
  $D_{eh}=n$ for $U > U'$ and  $D_{eh}=2n$ for $U < U'$.  
The former case corresponds to an exciton phase, and 
  the latter to a biexciton phase where 
  two electrons and two holes (i.e., two excitons) gather on a site.
Actually, exciton and biexciton insulating phases with the  
  aforementioned features have been obtained by using DMFT combined 
  with the exact diagonalization method.%
~\cite{Tomio05,TomioLT}

\begin{figure}[tb]
 \includegraphics[width=8cm,clip,keepaspectratio]{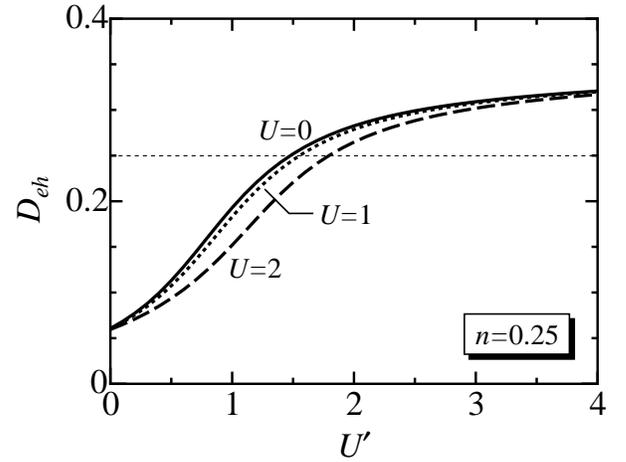}
\caption{%
The density of sites quadruply occupied by electrons and holes 
  as a function of $U'$ for $n=0.25$, $T=0.04$, and $\gamma=1$ 
  (these parameters are the same as in Fig.~\ref{Fig2}) 
  with fixed $U=0$, 1, and 2.
}%
\label{Fig3}
\end{figure}

In Fig.~\ref{Fig3}, the $U'$ dependence of $D_{eh}$ is shown 
  for $n=0.25$, $T=0.04$, and $\gamma=1$ 
  (corresponding to Fig.~\ref{Fig2}) with fixed $U=0$, 1, and 2.
For $U'=0$, the density of occupied sites becomes 
  $D_{eh} \simeq n^2=0.0625$, that is the value 
  in the noninteracting case, regardless the value of $U$.
With increasing $U'$, $D_{eh}$ increases monotonically.
This tendency clearly indicates the development of 
  the excitonic correlation toward the formation of local \e-h pairs. 
To detect trends more evidently, we drew the thin dotted line 
  ($D_{eh}=n=0.25$) as a loose guide for  
  a criterion of local \e-h pair formation. 
Here, of course, we should keep in mind that 
  the system is as metallic as ever, as seen in Fig.~\ref{Fig2}, 
  even if $D_{eh}$ is above the line.
One can find that the excitonic correlation is suppressed 
  by $U$ for $U \simeq U'$. 
In the large-$U'$ region where the effect of $U$ is suppressed, 
  $D_{eh}$ increases beyond 0.25 
  as a result of the enhancement of the biexcitonic correlation.
However, 
  $D_{eh}$ does not reach $2n=0.5$ corresponding to the biexciton phase. 
This is because the $t$-matrix approximation 
  cannot deal adequately with the four-body correlation,  
  irrespective of the LA. 
Hence problems of biexcitons and their condensation 
  are out of the scope of this paper.

\subsection{Renormalization factor}

Here we consider the case that 
  the electron and hole band masses are different ($\gamma \neq 1$).
Evaluating the renormalization factor, we examine 
  the properties of the normal phase in the case with mass difference.
We define the renormalization factor $Z_\alpha$ by
\begin{eqnarray}
\label{Z_alpha}
Z_\alpha^{-1} &=& 
1-\frac{d \Sigma_\alpha(\omega_n)}{d(i\omega_n)} \Big|_{i\omega_n \to 0}
\nonumber \\
&\simeq& 
1-\frac{\Im\left[
\Sigma_\alpha(i\pi T)-\Sigma_\alpha(-i\pi T)
\right]}{2\pi T}
\point
\end{eqnarray}
The quantity $Z_\alpha$ at $T \to 0$ 
  has generally several physical meanings characterizing 
  the Fermi liquid phase, e.g., 
  the weight of the quasiparticle coherent peak 
  in the density of states and 
  the jump at the Fermi wave number in the momentum 
  distribution function. 
Since the self-energy is momentum independent, 
  it also gives directly the effective mass 
  enhancement of quasiparticles, $m_\alpha/m_\alpha^*$.

\begin{figure}[tb]
 \includegraphics[width=8cm,clip,keepaspectratio]{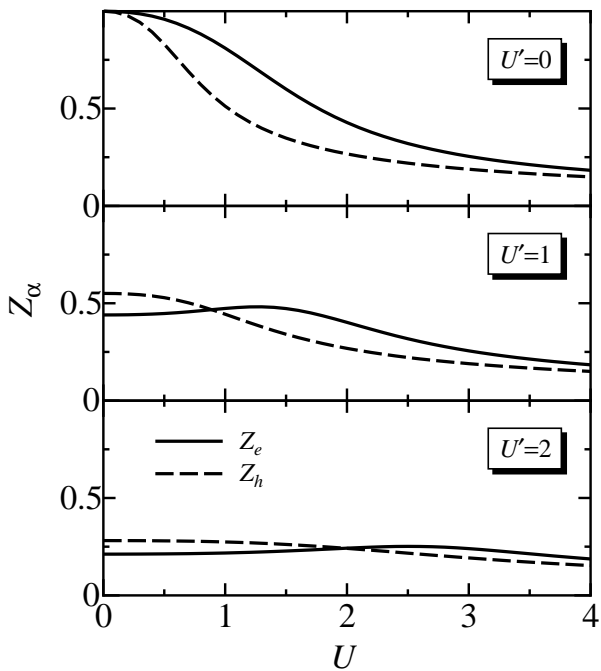}
\caption{%
The renormalization factor $Z_\alpha$ as a function of $U$
  for $n=0.5$, $T=0.05$, and $\gamma=0.5$
  with fixed $U'=0$, 1, and 2. 
The solid (dashed) curve denotes $Z_e$ ($Z_h$).
}%
\label{Fig4}
\end{figure}

In Fig.~\ref{Fig4}, we show the $U$ dependence of $Z_\alpha$ 
  for $n=0.5$ (quarter filling), $T=0.05$, and $\gamma=0.5$
  (the hole band mass is twice that of the electron) 
  with fixed $U'=0$, 1, and 2. 
For $U'=0$, 
  both $Z_e$ and $Z_h$ are monotonically decreasing functions 
  with respect to $U$. 
Because of the narrow hole band, 
  the effective repulsive interaction between holes $U/(2t_h)$ is 
  quite strong, leading to a large reduction of $Z_h$ (dashed curve). 
The presence of $U'$ ($U'=1$ and 2) reduces both $Z_e$ and $Z_h$. 
However, the behaviors of $Z_e$ and $Z_h$ for $U' \neq 0$ are  
  qualitatively different from those for $U'=0$: 
$Z_h$ decreases slowly and monotonically with increasing $U$, 
  while $Z_e$ has a slight hump (solid curve). 
The origin of this hump is the competition between 
  the effective repulsion $U/(2t_e)$ and 
  the effective \e-h attraction $U'/(t_e+t_h)$, 
  which raises the mobility of electrons.   
When $\gamma=0.5$, therefore, the hump of $Z_e$ appears at 
  $U \simeq 2t_e/(t_e+t_h) \times U'=2U'/(1+\gamma)\simeq 1.3$ and 2.7 
  for $U'=1$ and 2, respectively. 
For holes the influence of the competition between  
  $U/(2t_h)$ and $U'/(t_e+t_h)$ is quite weak,
  and $Z_h$ is almost constant in the region of 
  $U \lesssim 2\gamma U'/(1+\gamma) \simeq 0.7$ and 1.3 
  for $U'=1$ and 2, respectively. 
Here we point out that $Z_\alpha$ becomes insensitive to $U$ 
  for $U'=2$. When $U'=2$, the tendency to form bound \e-h pairs 
  begins to grow strongly, indicating very weak $U$ dependence.
This fact will also appear in the behavior of the transition temperature 
  discussed in the next section.

\section{Transition Temperature}

In this section, based on the self-consistent solution and 
  the condition defined in (\ref{Tc}), 
  we evaluate the transition temperature $T_c$ 
  from the normal phase to the \e-h pair condensed phase. 
By analyzing the $U'$ dependence of $T_c$ for various $U$ and $\gamma$, 
  the effects of the repulsive interaction 
  and the mass difference on the transition temperature are discussed.

\subsection{Effect of $U$}

First, we consider the effect of the repulsive interaction $U$ 
  on the transition temperature $T_c$. 
The calculations in this subsection are 
  restricted to the mass ratio $\gamma=1$.

\begin{figure}[tb]
 \includegraphics[width=8cm,clip,keepaspectratio]{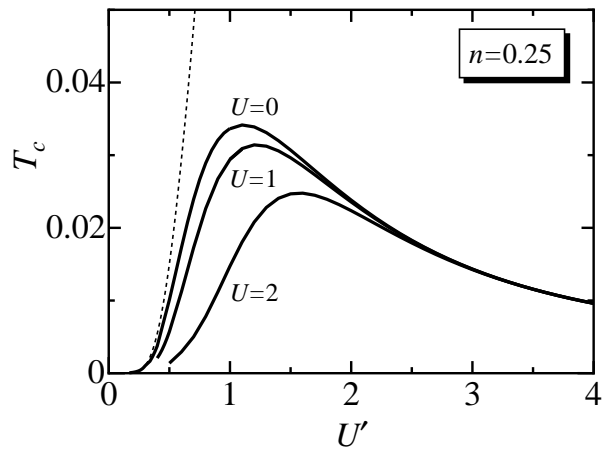}
\caption{%
The transition temperature $T_c$ as a function of $U'$ 
  for $n=0.25$ and $\gamma=1$ with fixed $U=0$, 1, and 2. 
The dotted line denotes the result from BCS theory. 
}%
\label{Fig5}
\end{figure}

Figure~\ref{Fig5} shows the $U'$ dependence of $T_c$ 
  for $n=0.25$ and $\gamma=1$ with fixed $U=0$, 1, and 2. 
The \e-h pair susceptibility in the normal phase diverges for 
  $T \searrow T_c$, which means that the \e-h pair condensed phase 
  is realized for $T < T_c$. 
For $U=0$, $T_c$ can be described well by the BCS result 
  (dotted curve) in the weak-coupling region ($U' \lesssim 0.3$), 
  that is, $T_c \propto \exp(-A/U')$ with a constant $A$.
With increasing $U'$, the solid curve deviates from the BCS result,  
  reaches a maximum at $U' \simeq 1$, and then decreases 
  as $1/U'$ for large $U'$. 
The $U'$ dependence of $T_c$ for large $U'$ is related to 
  the behavior of the BEC temperature of a lattice boson system%
~\cite{Nozieres85,Micnas90,Keller99} 
  with kinetic and potential energies of order $1/U'$.%
~\cite{Note1}
Obviously this result implies the BCS-BEC crossover, as expected. 
Note that the ladder terms for the dressed Green's function  
  is essential to extract the above successive crossover.%
~\cite{Fresard92}  
In the presence of $U$ ($U=1$ and 2), 
  one can see the reduction of $T_c$ for $U' \lesssim U$. 
This fact can be understood as a consequence of the suppression of 
  the excitonic correlation discussed 
  in Figs.~\ref{Fig2} and \ref{Fig3}. 
In contrast, $T_c$ is largely insensitive to the repulsion $U$  
  for $U' \gtrsim U$. 
This supports the validity of physical picture of strongly bound 
  \e-h pairs that behave almost like neutral bosons for $U' \gtrsim U$,  
  which is consistent with the analysis of the density of 
  occupied sites in Fig.~\ref{Fig3}.

\begin{figure}[tb]
 \includegraphics[width=8cm,clip,keepaspectratio]{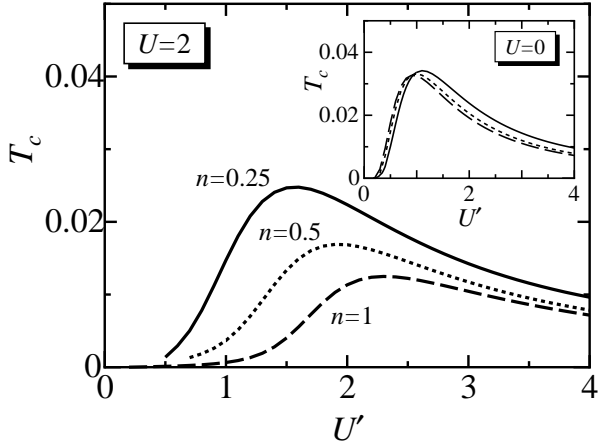}
\caption{%
The transition temperature $T_c$ as a function of $U'$ 
  for $\gamma=1$ and $U=2$ with fixed $n=0.25$, 0.5, and 1.0. 
The inset shows the case of $U=0$ for $n=0.25$ (solid line), 
  0.5 (dotted line), and 1.0 (dashed line). 
}%
\label{Fig6}
\end{figure}

Next, the effect of the repulsion $U$ on the transition temperature 
  is examined by changing the \e-h particle density. 
In Fig.~\ref{Fig6}, the $U'$ dependence of $T_c$ 
  with $\gamma=1$ and $U=2$ is shown for $n=0.25$, 0.5, and 1 
  (half filling).
The inset displays the corresponding $U'$ dependence of $T_c$ for $U=0$.
When $U=0$ (inset of Fig.~\ref{Fig6}), 
  the transition temperature $T_c$ increases slightly 
  in the weak-coupling region ($U' \ll 1$) 
  as the particle density $n$ is increased.
This behavior is quite reasonable for $U' \ll 1$ 
  from the following reason. 
Within the BCS theory the effective attractive interaction between 
  electrons and holes is roughly given by 
  $\rho_\alpha^0(\epsilon_F^{\alpha}) U'$ 
  where $\epsilon_F^\alpha$ is the Fermi energy 
  for an uncorrelated system [see Eq.~(\ref{TcBCS})]. 
Hence the effective attraction reaches maximum 
  when $n=1$, i.e., the band is half filled, in the case of 
  the semicircular density of states.  
The presence of $U$, however, completely changes this tendency.
As seen in the main figure of Fig.~\ref{Fig6} ($U=2$), 
  $T_c$ is sufficiently suppressed by $U$ for $U' \lesssim U$ 
  as $n$ approaches 1. 
When we simply consider the effect of $U$ in the BCS regime, 
  the effective interaction $\rho_\alpha^0(\epsilon_F^\alpha) U'$ 
  could be replaced by $\rho_\alpha(\epsilon_F^\alpha) U'$, 
  where $\rho_\alpha(\epsilon_F^\alpha)$ is 
  strongly renormalized by $U$, as seen in Fig.~\ref{Fig2}.
The renormalization becomes strong as $n$ approaches 1.%
~\cite{Menge91}
As a result, the effective interaction is reduced by $U$ 
  as $n \to 1$, leading to the suppression of $T_c$.  
Meanwhile, for large $U'$ the contribution of $U$ to $T_c$ is 
  very small regardless of the value of $n$.

\subsection{Effect of mass difference}

Finally, we investigate the effect of the mass difference $\gamma$ 
  on the transition temperature $T_c$.

\begin{figure}[tb]
 \includegraphics[width=8cm,clip,keepaspectratio]{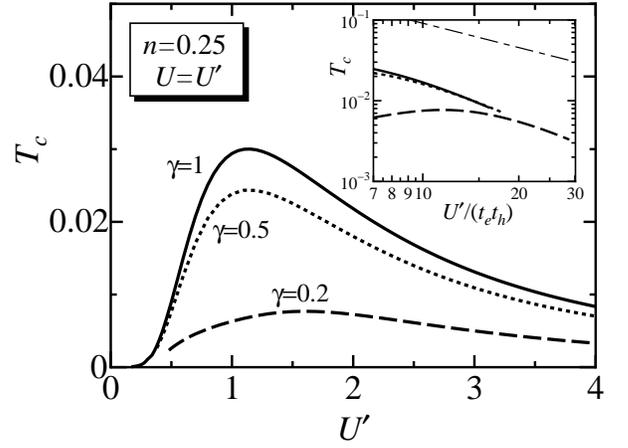}
\caption{%
The transition temperature $T_c$ as a function of $U'$ 
  for $n=0.25$ and $U=U'$ with fixed $\gamma=1$, 0.5, and 0.2.
In the inset, $T_c$ is shown as a function of $U'/(t_e t_h)$ for 
  $\gamma=1$ (solid line), 0.5 (dotted line), and 0.2 (dashed line), 
  where the decimal logarithm is taken for both axes.
The dot-dashed line represents the exact result~(\ref{TcBEC}) 
  in the limit of $U \to \infty$ and large $U'$. 
}%
\label{Fig7}
\end{figure}

We plot the $U'$ dependence of $T_c$ 
  for $n=0.25$ and $U=U'$ in Fig.~\ref{Fig7}, 
  with fixed $\gamma=1$, 0.5, and 0.2.
The mass difference has the effect of reducing 
  the transition temperature, 
  which is qualitatively consistent with the result of 
  the BCS-like pairing theory by Mizoo {\it et al.}%
~\cite{Mizoo05}
In addition to the reduction of $T_c$ in the weak-coupling BCS regime,  
  our result also involves the suppression of $T_c$ 
  in the strong-coupling BEC regime.
Such a behavior of $T_c$ can be understood analytically 
  in the weak- and strong-coupling limits as follows.

In the weak-coupling limit, the transition temperature
  can be estimated by using the Green's function 
  including only the Hartree term in its own self-energy, 
  which is given by
\begin{equation}
 \label{TcBCS}
T_c^{\rm BCS} = 
1.13 \sqrt{w_c^e w_c^h}
\,\exp \left(-\frac{t_e+t_h}
{2t_\alpha \rho_\alpha^0(\epsilon_F^\alpha) U'}\right)
\virg
\end{equation}
where $w_c^\alpha$ is a cutoff energy of order $\epsilon_F^\alpha$.
Note that $t_e \rho_e^0(\epsilon_F^e)=t_h \rho_h^0(\epsilon_F^h)$.
Equation~(\ref{TcBCS}) indicates that 
  the effective \e-h attractive interaction  
  is given by $t_\alpha \rho_\alpha^0(\epsilon_F^\alpha) U'$, 
  as mentioned above.
Moreover, another important point derived from Eq.~(\ref{TcBCS}) 
  is that the system is characterized by an energy scale $t_e+t_h$  
  determining the dimensionless effective coupling strength
  in the weak-coupling region.
In our model the quantity $(t_e+t_h)^{-1}$ is proportional to 
  $m_e m_h/(m_e+m_h)$, namely, the reduced mass.
Therefore, the transition to the \e-h pair condensed 
  phase in the weak-coupling region is related to 
  the {\it relative motion} 
  between electrons and holes, implying the BCS regime.
The $\gamma$ dependence of $T_c$ in the BCS regime can be 
  roughly evaluated from that of the cutoff energy $w_c^\alpha$, 
  where we should recall that the argument of the exponential 
  in Eq.~(\ref{TcBCS}) does not depend on the mass ratio $\gamma$ 
  when $t_e+t_h$ is taken as the energy unit.
The cutoff energy of holes $w_c^h$
  should be $\gamma w_c^e$ ($\sim \gamma t_e$) 
  since $\epsilon_F^h=\gamma\epsilon_F^e$ ($\sim \gamma t_e$).
Thus the coefficient of Eq.~(\ref{TcBCS}) becomes
  $\sqrt{w_c^e w_c^h} \sim \sqrt{\gamma}t_e$,   
  and so it is found that 
  $T_c$ in the BCS regime would be approximately 
  proportional to $\sqrt{\gamma}/(1+\gamma)$.

On the other hand, in the strong-coupling limit,  
  the model~(\ref{H}) can be mapped onto a single-band attractive 
  Hubbard model with the interaction $-U'$ and the hopping
  $t_\alpha$ at $U=\infty$. 
For large $U'$,  
  this attractive Hubbard model can map to a hard core Boson model 
  with the kinetic energy $2t_e t_h/U'$ and the potential energy 
  $(t_e^2+t_h^2)/U'$.  
Using the standard mean-field theory, which becomes exact 
  in the limit of the large coordination number of the lattice,  
  we can obtain the BEC temperature in the limit of $U \to \infty$ 
  and large $U'$ as~\cite{Note2}
\begin{equation}
 \label{TcBEC}
T_c^{\rm BEC} =
\frac{2t_e t_h}{U'} \frac{2n-1}{\ln\left[n/(1-n)\right]}
\point
\end{equation}
This expression indicates that the system is 
  characterized by an energy scale $t_e t_h/(t_e+t_h)$ 
  determining the dimensionless effective coupling strength
  in the strong-coupling region, 
  which is related to the {\it motion of the center of mass} 
  since $(t_e+t_h)/(t_e t_h) \propto m_e+m_h$. 
The $U'$ dependence of $T_c$ in the strong-coupling region 
  of Fig.~\ref{Fig7} can be described asymptotically 
  in terms of $t_e t_h/U'$. In the inset of Fig.~\ref{Fig7}, 
  the transition temperature is shown together with the result 
  of Eq.~(\ref{TcBEC}), where $U'/(t_e t_h)$ is taken as the $x$ axis 
  and the decimal logarithm is taken for both axes.
The transition temperature $T_c$ for various mass ratio $\gamma$ tends to 
  behave linearly with the slope of $-1$ (i.e., $T_c \propto 1/U'$)
  and merges in the large-$U'$ region, implying the BEC regime.
As seen from Eq.~(\ref{TcBEC}),  
  the $\gamma$ dependence of $T_c$ in the BEC regime is 
  given by $T_c \propto t_e t_h = \gamma/(1+\gamma)^2$.
Therefore, it is confirmed that for the fixed $U'$ and $n$ 
  the transition temperature is suppressed 
  by the \e-h mass difference in the BEC regime,  
  as well as in the BCS regime.

Viewed in this way, the characteristic energy scale 
  and the \e-h mass ratio dependence of the transition temperature
  become clear in both the BCS and BEC regimes. 
We especially emphasize the point that 
  the excitonic BCS-BEC crossover can be marked 
  by the change of the characteristic energy scale 
  from $t_e+t_h$ to $t_e t_h/(t_e+t_h)$, 
  which is not necessarily evident in the standard scenario%
\cite{Randeria}
  of the BCS-BEC crossover.

\section{Concluding Remarks}

In this paper, 
  by applying the self-consistent $t$-matrix approximation 
  with the LA to the \e-h two-band Hubbard model, 
  we have discussed the properties of the normal phase and 
  the transition temperature to the \e-h pair condensed phase.  
In the analysis, the effects of the repulsive interaction 
  between like particles and the mass \e-h difference 
  have been given special attention.

From the behavior of the calculated physical quantities,
  we have found that the development of the excitonic
  correlation induced by the \e-h attractive interaction   
  is suppressed in the presence of the repulsion between 
  like particles and also the mass difference.
In particular, it is noteworthy that 
  the competition between the effective interactions 
  $U/(2t_\alpha)$ and $U'/(t_e+t_h)$ play the key roles in 
  controlling the formation of excitonic bound pairs 
  in the normal phase.
This is a remarkable feature of the two-band \e-h system.

The behavior of the transition temperature has shown that 
  the present SCTMA combined with the LA 
  can describe well the excitonic BCS-BEC crossover 
  at finite temperatures. 
In the BCS regime of weak \e-h attraction, 
  the repulsive interaction sufficiently suppresses 
  the transition temperature for $U > U'$. 
In contrast, it does not affect the transition temperature 
  for $U < U'$ in the BEC regime of strong \e-h attraction.  
These results imply that  
  the competition between the effective repulsion and 
  attraction plays an important role in understanding 
  the properties of the \e-h pair condensed phase 
  and the crossover behavior also below $T_c$.
Furthermore, we have found that 
  in the whole BCS-BEC regime the \e-h mass difference 
  leads to large suppression of the transition temperature.
This analysis in the case of $\gamma \neq 1$ has allowed us 
  to capture the BCS-BEC crossover as 
  the change of the characteristic energy scale from $t_e+t_h$ 
  to $t_e t_h/(t_e+t_h)$, where the former is related to 
  the relative motion 
  and the latter to the motion of the center of mass.

Let us now discuss limitations of the present SCTMA.
The $n$ dependence of $T_c$
  in the large-$U'$ region is opposite to that of 
  the exact result in the limit of $U \to \infty$ and large $U'$ 
  given by Eq.~(\ref{TcBEC}).
The result of Fig.~\ref{Fig6} suggests an decreasing function of $n$
  while the exact result is an increasing function of $n$.
It is not clear at present why these are not consistent 
  even qualitatively. 
At least we should remind ourselves that the $t$-matrix approximation 
  is valid for the low-density limit.

Another difficulty is that the SCTMA 
  cannot describe insulating states without any symmetry breaking,%
~\cite{Menge91,Keller99}
  such as the exciton and biexciton phases.  
Here we should remark that the SCTMA for 
  the single-band attractive Hubbard model%
~\cite{Keller99} 
  does not yield the pseudogap behavior 
  (or the pairing transition%
~\cite{dosSantos94,Capone02,Keller02,Toschi05})
  as seen in high-$T_c$ superconductors.
The pseudogap phase in the context of superconductors
  is deeply related to the exciton phase in \e-h systems 
  where incoherent \e-h bound pairs 
  (which do not condense but are insulating) are formed. 
In our model, the DMFT combined with 
  the exact diagonalization method%
~\cite{Tomio05,TomioLT} 
  indicates the appearance of 
  exciton and biexciton insulating phases that already show up 
  for $U' \simeq 2$.
Nevertheless, it is still notable that even within the SCTMA 
  one can see the tendency to form \e-h bound pairs 
  above $T_c$ by calculating the density of states and 
  the density of occupied sites. 
However, a unified theory for 
  the exciton Mott transition and the excitonic condensation 
  would require nonperturbative approaches 
  such as the full DMFT 
  (Refs.~\onlinecite{Tomio05,TomioLT}, and \onlinecite{Toschi05}) 
  and Monte Carlo methods.%
~\cite{DePalo02,Koch03}

The long-range part of the Coulomb interaction was not 
  considered here. 
It may become crucial particularly in the BEC regime, 
  which motivates future work.

As has been argued in Refs.%
~\onlinecite{InagakiSSC} and \onlinecite{Inagaki02}, 
  clarifying the BCS-BEC crossover problem from the optical response 
  attracts great interest and is important from not only 
  the theoretical but also the experimental aspect.  
For this purpose, the present SCTMA will be extended to the 
  \e-h pair condensed phase below $T_c$. 
The analysis for the properties of the condensed phase and 
  optical response will be reported soon in a forthcoming presentation.%
~\cite{TomioFCP}

\begin{acknowledgments}
The authors would like to thank K.~Asano, A.~Tsuruta, T.~Tohyama, 
and A.~Koga for useful discussions and comments. 
We also thank P.~Huai for a critical reading of this paper. 
The computation in this work has been done using the facilities of 
the Supercomputer Center, Institute for Solid State Physics, 
University of Tokyo. This work has been supported by 
Core Research for Evolutional Science and Technology (CREST), 
Japan Science and Technology Agency (JST).
\end{acknowledgments}



\begin{thebibliography}{99} 
\bibitem{Moskalenko00}
S.~A.~Moskalenko and D.~W.~Snoke, 
{\it Bose-Einstein Condensation of Excitons and Biexcitons} 
(Cambridge University Press, Cambridge, U.K., 2000).
\bibitem{Haug84}
H.~Haug and S.~Schmitt-Rink, 
Prog.\ Quantum Electron.\ {\bf 9}, 3 (1984). 
\bibitem{Zimmermann}
R.~Zimmermann, 
{\it Many-Particle Theory of Highly Excited Semiconductors} 
(Teubner, Leipzig, 1988).
\bibitem{Moskalenko62}
S.~A.~Moskalenko, 
Fiz.\ Tverd.\ Tela (Leningrad) {\bf 4}, 276 (1962) 
[Sov.\ Phys.\ Solid State {\bf 4}, 199 (1962)]. 
\bibitem{Blatt62}
J.~M.~Blatt, K.~W.~B\"oer, and W.~Brandt,
Phys.\ Rev.\ {\bf 126}, 1691 (1962).
\bibitem{Keldysh65}
L.~V.~Keldysh and Y.~V.~Kopaev, 
Fiz.\ Tverd.\ Tela (Leningrad) {\bf 6}, 2791 (1964) 
[Sov.\ Phys.\ Solid State {\bf 6}, 2219 (1965)]. 
\bibitem{Jerome67}
D.~J\'erome, T.~M.~Rice, and W.~Kohn, 
Phys.\ Rev.\ {\bf 158}, 462 (1967).
\bibitem{Snoke03}
D.~W.~Snoke, 
Phys.\ Status Solidi B {\bf 238}, 389 (2003).
\bibitem{Snoke90}
D.~W.~Snoke, J.~P.~Wolfe, and A.~Mysyrowicz, 
Phys.\ Rev.\ Lett.\ {\bf 64}, 2543 (1990); 
Phys.\ Rev.\ B {\bf 41}, 11171 (1990). 
\bibitem{Lin93}
J.~L.~Lin and J.~P.~Wolfe, 
Phys.\ Rev.\ Lett.\ {\bf 71}, 1222 (1993).
\bibitem{Gonokami02}
M.~Kuwata-Gonokami, R.~Shimano, and A.~Mysyrowicz,  
J.\ Phys.\ Soc.\ Jpn.\ {\bf 71}, 1257 (2002).
\bibitem{Vasilev04}
P.~P.~Vasil'ev, 
Phys.\ Status Solidi B {\bf 241}, 1251 (2004).
\bibitem{Comte82}
C.~Comte and P.~Nozi\`eres, 
J.\ Phys.\ (France) {\bf 43}, 1069 (1982); 
P.~Nozi\`eres and C.~Comte, 
{\it ibid.}\ {\bf 43}, 1083 (1982).
\bibitem{InagakiSSC}
T.~J.~Inagaki, M.~Aihara, and A.~Takahashi, 
Solid State Commun.\ {\bf 115}, 645 (2000).
\bibitem{Inagaki02}
T.~J.~Inagaki and M.~Aihara,
Phys.\ Rev.\ B {\bf 65}, 205204 (2002).
\bibitem{Littlewood04}
P.~B.~Littlewood, P.~R.~Eastham, J.~M.~J.~Keeling, 
F.~M.~Marchetti, B.~D.~Simons, and M.~H.~Szymanska, 
J.\ Phys.: Condens.\ Matter {\bf 16}, S3597 (2004).
\bibitem{Nozieres85}
P.~Nozi\`eres and S.~Schmitt-Rink, 
J.\ Low Temp.\ Phys.\ {\bf 59}, 195 (1985). 
\bibitem{Micnas90}
R.~Micnas, J.~Ranninger, and S.~Robaszkiewicz, 
Rev.\ Mod.\ Phys.\ {\bf 62}, 113 (1990). 
\bibitem{Randeria}
M.~Randeria, 
in {\it Bose-Einstein Condensation}, 
edited by A.~Griffin, D.~W.~Snoke, and S.~Stringari  
(Cambridge University Press, Cambridge, U.K., 1995), p.~355.
\bibitem{Ohashi02}
Y.~Ohashi and A.~Griffin, 
Phys.\ Rev.\ Lett.\ {\bf 89}, 130402 (2002); 
Phys.\ Rev.\ A {\bf 67}, 033603 (2003).
\bibitem{Kopaev66}
Y.~V.~Kopaev, 
Fiz.\ Tverd.\ Tela (Leningrad) {\bf 8}, 223 (1966)  
[Sov.\ Phys.\ Solid State {\bf 8}, 175 (1966)]. 
\bibitem{Zittartz67}
J.~Zittartz, 
Phys.\ Rev.\ {\bf 162}, 752 (1967).
\bibitem{Mizoo05}
K.~Mizoo, T.~J.~Inagaki, Y.~Ueshima, and M.~Aihara,
J.\ Phys.\ Soc.\ Jpn.\ {\bf 74}, 1745 (2005).
\bibitem{Fresard92}
R.~Fr\`esard, B.~Glaser, and P.~W\"olfle, 
J.\ Phys.:\ Condens.\ Matter {\bf 4}, 8565 (1992).
\bibitem{Haussmann}
R.~Haussmann, 
Z.\ Phys.\ B: Condens.\ Matter {\bf 91}, 291 (1993); 
Phys.\ Rev.\ B {\bf 49}, 12975 (1994).
\bibitem{Keller99}
M.~Keller, W.~Metzner, and U.~Schollw\"{o}ck,
Phys.\ Rev.\ B {\bf 60}, 3499 (1999).
\bibitem{Letz98}
M.~Letz and R.~J.~Gooding,
J.\ Phys.:\ Condens.\ Matter {\bf 10}, 6931 (1998).
\bibitem{Georges96}
A.~Georges, G~.Kotliar, W.~Krauth, and M.~J.~Rozenberg, 
Rev.\ Mod.\ Phys.\ {\bf 68}, 13 (1996). 
\bibitem{Note0}
We restrict our consideration to a pure $s$-wave pairing, 
  which is valid for low \e-h density or for $U=0$ and $\infty$.
Of course, the possibility of the state having 
  other pairing symmetry (such as $d$ wave) is not excluded, 
  especially in the case near half filling and with finite $U$.
To examine this, however, 
  the momentum dependence of the self-energy is required.
\bibitem{Koga02}
A.~Koga, Y.~Imai, and N.~Kawakami, 
Phys.\ Rev.\ B {\bf 66}, 165107 (2002).
\bibitem{Tomio05}
Y.~Tomio and T.~Ogawa, 
J.\ Lumin.\ {\bf 112}, 220 (2005).
\bibitem{Menge91}
B.~Menge and E.~M\"{u}ller-Hartmann,
Z.\ Phys.\ B: Condens. Matter {\bf 82}, 237 (1991).
\bibitem{TomioLT}
Y.~Tomio and T.~Ogawa, 
in Proceedings of the International Conference 
on Low Temperature Physics, Orlando, Florida, 2005, 
edited by Y. Takano (unpublished).
\bibitem{Note1}
$T_c \propto 1/U'$ in the lattice model corresponds to 
  the BEC temperature in the continuum model where  
$T_c$ saturates and does not depend on the interaction strength 
  in the BEC regime; 
  see, for example, Ref.~\onlinecite{Randeria}.
\bibitem{Note2}
The expression of $T_c^{\rm BEC}$ 
  corresponding to the case of $t_e=t_h$ 
  is seen in Refs.~\onlinecite{Micnas90} and \onlinecite{Keller02}.  
\bibitem{Keller02}
M.~Keller, W.~Metzner, and U.~Schollw\"ock, 
J.\ Low\ Temp.\ Phys.\ {\bf 126}, 961 (2002).
\bibitem{dosSantos94}
R.~R.~dos Santos, 
Phys.\ Rev.\ B {\bf 50}, R635 (1994).
\bibitem{Capone02}
M.~Capone, C.~Castellani, and M.~Grilli, 
Phys.\ Rev.\ Lett.\ {\bf 88}, 126403 (2002).
\bibitem{Toschi05}
A.~Toschi, P.~Barone, M.~Capone, and C.~Castellani, 
New J.\ Phys.\ {\bf 7}, 7 (2005).
\bibitem{DePalo02}
S.~De Palo, F.~Rapisarda, and G.~Senatore, 
Phys.\ Rev.\ Lett.\ {\bf 88}, 206401 (2002).
\bibitem{Koch03}
S.~W.~Koch, W.~Hoyer, M.~Kira, and V.~S.~Filinov,
Phys.\ Status Solidi B {\bf 238}, 404 (2003).
\bibitem{TomioFCP}
Y.~Tomio and T.~Ogawa (unpublished).

\end{thebibliography}
\end{document}